\documentclass[12pt]{iopart}

\usepackage{latexsym}
\usepackage{amssymb}
\usepackage{graphicx}                            %  include figure files
\usepackage{epsfig}                              %  files .eps
\usepackage{dcolumn}                             %  align tble columns on decimal points
\usepackage{color}                               %  color fonts
\usepackage{ulem}                                %  strike out (sout)
\usepackage{bm}                                  %  bold math
\usepackage[colorlinks=true,dvipdfm]{hyperref}   %  to include internet links
\newcommand{\Epsilon}{\mathcal{E}}

\begin{document}

% TITLE, AUTHORS AND ABSTRACT
%%%%%%%%%%%%%%%%%%%%%%%%%%%%%%%%%%%%%%%%%%%%%%%%%%%%%%%%%%%%%%%%%%%%%%%%%%%%%%%%%%%%%%%%%%%%%%
\title{Energy dispersion in radiation pressure accelerated ion beams}

\author{M.~Grech}
\ead{mickael.grech@gmail.com}
\address{Max-Planck-Institute for the Physics of Complex Systems, D-01187 Dresden, Germany}

\author{S.~Skupin}
\address{Max-Planck-Institute for the Physics of Complex Systems, D-01187 Dresden, Germany}
\address{Institute of Condensed Matter Theory and Solid State Optics, Friedrich-Schiller-University Jena, D-07743 Jena, Germany}

\author{A.~Diaw}
\address{Centre de Physique Th\'{e}orique, Ecole Polytechnique, F-91128 Palaiseau, France}

\author{T.~Schlegel}
\address{Helmholtz Institute Jena, D-07743 Jena, Germany}

\author{V.~T.~Tikhonchuk}
\address{Centre Lasers Intenses et Applications, F-33405 Talence, France}

\begin{abstract}
We address the problem of energy dispersion of radiation pressure accelerated (RPA) ion beams emerging from a thin (solid) target. Two different acceleration schemes, namely phase-stable acceleration and multi-stage acceleration, are considered by means of analytical modelling and one-dimensional particle-in-cell simulations. Our investigations offer a deeper understanding of RPA and allow us to derive some guidelines for generating monoenergetic ion beams.
\end{abstract}

\maketitle
%%%%%%%%%%%%%%%%%%%%%%%%%%%%%%%%%%%%%%%%%%%%%%%%%%%%%%%%%%%%%%%%%%%%%%%%%%%%%%%%%%%%%%%%%%%%%%

%%%%%%%%%%%%%%%%
% INTRODUCTION %
%%%%%%%%%%%%%%%%    
\section{Introduction}\label{sec1}

Interaction of ultra-intense laser pulses with thin foils offers interesting possibilities to generate energetic charged particles. So-called radiation pressure acceleration (RPA) of ion bunches has recently attracted a lot of interest as it may provide an efficient way to generate intense quasi-monoenergetic ion beams. In contrast to target normal sheath acceleration (TNSA)~\cite{wilks_POP_2001,mora_PRL_2003}, where ions are accelerated from the target rear surface (the front surface being the one irradiated by the laser pulse) in the electrostatic field built up by the laser-created hot electrons, RPA of ion beams relies on the efficient momentum transfer from laser photons to ions in a thin dense target, which reflects the incident laser pulse. RPA turns out to be a very efficient way to accelerate quasi-neutral ion-electron bunches up to potentially relativistic velocities while keeping energy dispersion small.

The idea of accelerating (macroscopic) objects by use of laser radiation pressure was initially discussed by Marx~\cite{marx_NATURE_1966} as a possible path toward interstellar space travel. Its application to efficient ion acceleration was first proposed in Ref.~\cite{esirkepov_PRL_2004}, where the authors show that, in order to observe efficient RPA with linearly polarized light, ultra-high laser intensities ($I_L \gtrsim 10^{23}~{\rm W/cm^2}$) are required. For lower intensities indeed, the strong laser-induced electron heating makes target expansion the dominant acceleration mechanism. In 2005, Macchi~{\it et al.} demonstrated that using circularly polarized (CP) laser light strongly reduces electron heating thus allowing RPA to operate efficiently at lower intensities ($I_L \gtrsim 10^{20}~{\rm W/cm^2}$)~\cite{macchi_PRL_2005}. After that, different teams have discussed independently the possibility to create quasi-monoenergetic ion beams by irradiating a thin target with an intense CP laser pulse~\cite{zhang_POP_2007,klimo_PRSTAB_2008,robinson_NJP_2008,yan_PRL_2008}. Many studies have followed, some proposing and/or revisiting different acceleration models or discussing the optimal laser-target parameters through one-dimensional (1D) and two-dimensional (2D) particle-in-cell (PIC) simulations~\cite{rykovanov_NJP_2008, tripathi_PPCF_2009, macchi_PRL_2009, qiao_PRL_2009}. Multi-dimensional effects on the stability of the accelerated foil and their potential capability to improve RPA have also been discussed in Refs.~\cite{pegoraro_PRL_2007} and~\cite{bulanov_PRL_2010}. Finally, first experimental confirmation of RPA has recently been claimed~\cite{henig_PRL_2009}.

While the basic mechanisms of RPA are now well understood, the control of the energy dispersion in RPA ion beams has not been fully addressed. This paper is intended to provide the reader with guidelines how to achieve this control. To do so, we first recall the basic modelling of RPA of a thin foil and provide scaling laws concerning the maximum ion energy that can be reached as a function of the laser intensity or power (Sec.~\ref{sec2}). Beyond this simple (macroscopic) modelling, we discuss the details of RPA of a thin foil as following from two complementary processes. For the thinnest foils, acceleration proceeds in the so-called phase-stable way~\cite{yan_PRL_2008}. For thick enough targets, acceleration occurs as a multi-stage process~\cite{zhang_POP_2007,klimo_PRSTAB_2008}. In Sec.~\ref{sec3}, we propose some refined models for both acceleration processes and extract the main requirements for controlling the resulting ion energy dispersion. These analytical findings are compared to numerical PIC simulations in Sec.~\ref{sec4}. Finally, we present our conclusions in Sec.~\ref{sec5}.

%%%%%%%%%%%%%%%%%%%%%
% BASIC  MODELLING  %
%%%%%%%%%%%%%%%%%%%%%
\section{Basic Modelling}\label{sec2}

In this paper, all quantities are normalized to laser and electron related units. Times and distances are normalized to the incident laser frequency $\omega_L$ and wave-number $k_L=\omega_L/c$, respectively, and velocities are normalized to the light velocity $c$. Electric charges and masses are normalized to the electron charge $e$ and mass $m_e$, respectively. Densities are normalized to the critical density at the considered laser wavelength $\lambda_L = 2\pi/k_L$: $n_c = \epsilon_0\,m_e\,\omega_L^2/e^2$, where $\epsilon_0$ is the permittivity of free space. Electric fields are normalized to the Compton field $E_C = m_e\,c\,\omega_L/e$. Furthermore, we consider a CP laser pulse and introduce the incident laser pulse vector potential:
\begin{eqnarray}
{\bf A}_L(t,x)=\frac{a_L(t)}{\sqrt{2}}\,\Big[\cos(t-x)\,{\bf \hat{y}} + \sin(t-x)\,{\bf \hat{z}} \Big]\,,
\end{eqnarray}
where ${\bf \hat{y}}$ and ${\bf \hat{z}}$ denote the two directions transverse to the laser propagation direction ${\bf \hat{x}}$.

\subsection{Macroscopic approach: the light-sail model}\label{sec2.1}

A straightforward and elegant way to derive the energy ion gain during the acceleration of a thin target by laser radiation pressure is to consider the accelerated layer as a quasi-neutral light sail reflecting the incident laser pulse~\cite{esirkepov_PRL_2004}. Acceleration then follows from momentum transfer from the laser photons to the ions. Assuming that all ions in the target have the same velocity $v_i^{(l)}$ in the laboratory-frame, the equation of motion for the target is obtained by equating the photon momentum flux to the ion momentum flux, which follows from the target acceleration $(n_{i0}\,d_{0})\,dp_i^{(l)}/dt$, where $p_i^{(l)}$ is the ion momentum in units of $m_e\,c$, $n_{i0}$ is the initial target ion density and $d_0$ its thickness. For arbitrary target velocities, two effects -- the reduction of the photon flux on the target and the Doppler shift lowering of the photon momenta in the target-frame -- must be taken into account. Considering total reflection of the laser pulse in the target-frame one obtains:
\begin{eqnarray}\label{eq_motion_foil}
(n_{i0}\,d_0)\,\frac{d}{dt}p_i^{(l)} = a_L^2\big[t-x_i(t)\big]\,\gamma_i^2\,\left(1-v_i^{(l)}\right)^2\,,
\end{eqnarray}  
where $a_L^2(t)$ is the normalized laser intensity, $x_i(t)$ is the time-dependent position of the target moving with the velocity $v_i^{(l)}$, and $\gamma_i=(1-v_i^{(l)2})^{-1/2}$ is the associated Lorentz factor. The solution of Eq.~(\ref{eq_motion_foil}) has been derived in Ref.~\cite{robinson_NJP_2008} for a laser pulse with a step-like temporal profile [$a_L^2(t)=a_0^2$ for $t>0$ and $a_L^2(t)=0$ otherwise], with the maximum laser field amplitude $a_0$:
\begin{eqnarray}\label{eq_pi_robinson}
p_i^{(l)}(t) = m_i\,\left[ {\rm sinh} \phi - \big(4\,{\rm sinh} \phi\big)^{-1} \right]\,,
\end{eqnarray} 
where $\phi = (1/3)\,{\rm sinh^{-1}} \big(3\,a_0^2\,t/(n_{i0}\,m_i\,d_0)+2\big)$ and $m_i$ is the ion mass. 

At this point, we stress that, in the derivation of Eq.~(\ref{eq_motion_foil}), (i) the electron momenta have been neglected, and (ii) the target is assumed to be thin enough to be accelerated as a whole, quasi-neutral bunch, but thick enough to support the laser radiation pressure. Both assumptions are discussed in more details in Sec.~\ref{sec3}.

\subsection{Scaling laws for the ion energy}\label{sec2.2}

Let us now derive some scaling laws for the ion energy $\Epsilon_i=(\gamma_i-1)\,m_i$ (in units of $m_e\,c^2$). First, Eq.~(\ref{eq_pi_robinson}) can be simplified in the limit of non-relativistic ions ($p_i^{(l)} \ll 1$), leading to $p_i^{(l)}(t) \sim a_0^2\,t/(n_{i0}\,d_0)$ and ion energies:
\begin{eqnarray}\label{eq_ionenergy_nonrelat}
\Epsilon_i \sim \frac{m_i}{2}\,\left(\frac{a_0^2\,t}{m_i\,n_{i0}\,d_0}\right)^2.
\end{eqnarray}
This result suggests that the ion energy scales with the square of the laser intensity $I_L = a_0^2$ (in units of $c^3\,m_e\,n_c/2$), and more precisely with the square of the laser fluence $\phi_L = \int_0^t a_L^2(t)\,dt$ [in units of $c^3\,m_e\,n_c/(2\,\omega_L)$]. However, this scaling applies only for sufficiently short laser pulses. Indeed, if the laser pulse is long enough for the target to travel over a distance larger than the laser Rayleigh length $L_R \sim w_L^2$ ($w_L$ is the transverse size of the laser focal spot), diffraction of the laser pulse must be accounted for. It sets in after a time $t \sim \sqrt{2\,m_i\,n_{i0}\,d_0}\,w_L/a_0$, so that the final ion energy is limited to $\Epsilon_i \sim a_0^2\,w_L^2/(n_{i0}\,d_0)$. For a sufficiently long laser pulse, the ion energy thus scales (only) linearly with the laser intensity. More precisely, one can introduce the normalized laser power onto the target, $P_L \sim a_0^2\,w_L^2$ [in units of $m_e\,n_c\,c^3/(2\,k_L^2)$], and we obtain that the final ion energy scales linearly with the laser power. 

These estimates suggest that relativistic ion energies can be reached for laser powers $P_{L} > m_i\,n_{i0}\,d_0$. Considering characteristic diamond-like carbon (DLC) targets ($m_i = 12 \times 1836$, $n_i = 60$ at $\lambda_L = 1~{\rm \mu m}$) with thickness $d/\lambda_L = 10^{-2}$~\cite{liechtenstein_NIMA_2006}, this corresponds to a power $P_L > 3.5 \times 10^4$, typically $\simeq 30$~TW. Hence, current laser facilities could be considered for relativistic ion generation. 

Equation~(\ref{eq_pi_robinson}) can also be simplified in the limit of ultra-relativistic ions $p_i^{(l)} \gg 1$:
\begin{eqnarray}\label{eq_ionenergy_ultrarelat}
\Epsilon_i \sim p_i^{(l)} \sim \frac{m_i}{2}\,\left(\frac{6\,a_0^2\,t}{m_i\,n_{i0}\,d_0}\right)^{1/3}\,.
\end{eqnarray}
The ion energy thus increases as $t^{1/3}$. This characteristic evolution has been first observed in the original paper by Esirkepov {\it et al.}~\cite{esirkepov_PRL_2004}. In the ultra-relativistic regime, $v_i^{(l)} \sim 1$, and diffraction sets in after a time $t \sim w_L^2$. The final ion energy then scales as the power $1/3$ of both the laser intensity and power. This is mainly because of photon red-shifting and the reduced photon flux onto the target due to its relativistic recoil.

%%%%%%%%%%%%%%%%%%%%%%%
% TWO REGIMES OF RPA  %
%%%%%%%%%%%%%%%%%%%%%%%
\section{Two RPA regimes}\label{sec3}

A deeper insight into RPA of thin targets requires to investigate more closely the structure of the accelerated target. When exposed to an intense laser pulse, the target electrons are pushed forward into the target by the laser ponderomotive force thus forming a compressed electron layer (CEL) at the laser front. The formation of this layer occurs on a very short time as it involves only electron motion. Its characteristic position $l_c$ before ion motion sets in can be easily derived from equating the electrostatic pressure $(Z\,n_{i0}\,l_c)^2/2$ to the radiation pressure $a_0^2$ on the CEL. Obviously, maintaining the target integrity requires $l_c$ to be smaller than the target thickness $d_0$. It is thus quite natural to introduce the normalized parameter $\xi = Z\,n_{i0}\,d_0/(\sqrt{2}\,a_0)$. For $\xi < 1$, the radiation pressure is so strong that it cannot be balanced by the electrostatic pressure inside the target. All electrons are expelled from the target which than undergoes Coulomb explosion. This regime of interaction has been studied in Refs.~\cite{kulagin_PRL_2007,grech_NJP_2009,grech_NIMA_2010}, where authors have considered its applications to both electron or ion acceleration. It is however not suitable for efficient RPA of thin foils, which requires $\xi \ge 1$. 

In what follows, we analyze two regimes of RPA of thin foils for parameter values $\xi \sim 1$ or $\xi \gg 1$.

\subsection{Phase-stable acceleration of thin targets}\label{sec3.1}

In the regime where $\xi \sim 1$, all electrons are piled-up at the rear-side of the target and ion acceleration proceeds in the so-called phase-stable regime~\cite{yan_PRL_2008} (also referred to as self-organized double-layer acceleration~\cite{tripathi_PPCF_2009} or coherent acceleration of ions by laser~\cite{tajima_RAST_2009}). In this specific regime, ion acceleration can be described by considering that all electrons form a compressed layer at the target rear-side. For the sake of simplicity, we assume that the electron density in the CEL is constant and homogeneous: $n_{ec} \sim Z\,n_{i0}\,d_0/l_e$, where $l_e<d_0$ is the CEL thickness. At this point, we stress that simple considerations on the balance between electrostatic and radiation pressures do not allow us to derive this thickness $l_e$. For $\xi = 1$, $l_e$ would indeed shrink to $0$, which is prevented by the electron pressure that is not included in our model. 

As ions initially located outside the CEL are accelerated in an electrostatic field, which increases linearly in space, they are, {\it a priori}, not of interest for efficient generation of quasi-monoenergetic ion bunches. We should thus focus our attention on the ions located in the CEL which undergo acceleration in the monotonically decreasing field:
\begin{eqnarray}
E_x(t,x) = E_0(t)\,\frac{d(t)-x}{l_e}\,,
\end{eqnarray}
where $d(t)$ is the position of the target rear-side at time $t$, and the maximum electrostatic field $E_0(t)$ can be derived from the equilibrium condition of the CEL (in the frame moving with the target rear-side):
\begin{eqnarray}
\frac{1}{2}\,(Z\,n_{i0}\,l_c)\,E_0(t) = a_0^2\,\gamma_i^2\,\left(1-v_i^{(l)}\right)^2,
\end{eqnarray}
where $v_i^{(l)}$ is the velocity (in the laboratory-frame) of the CEL, i.e. of ions accelerated in a phase-stable way, and $\gamma_i = (1-v_i^{(l)2})^{-1/2}$ is the associated Lorentz factor. From this we obtain:
\begin{eqnarray}
E_0(t) = \sqrt{2}\,a_0\,\gamma_i^2\,\left(1-v_i^{(l)}\right)^2\,.
\end{eqnarray}
Note that in this regime, where electrons are piled-up at the rear-side of the target, the accelerating field $E_0(t)$ does not depend on the CEL thickness $l_e$. A similar feature was discussed in Ref.~\cite{macchi_PRL_2009}.

The governing equation for the mean ion momentum can be easily derived by considering that, in the phase-stable regime, ions are accelerated, in the average, by an electrostatic field $E_0(t)/2$. We obtain:
\begin{eqnarray}
\frac{d}{dt}p_i^{(l)} = \frac{Z\,a_0}{\sqrt{2}}\,\gamma_i^2\,\left(1-v_i^{(l)}\right)^2,
\end{eqnarray}
which is nothing but Eq.~(\ref{eq_motion_foil}) derived in the macroscopic model for $\xi \sim 1$. 

As for the ion motion around the mean velocity $v_i^{(l)}$, it can be described as in Ref.~\cite{yan_PRL_2008}. Denoting $\chi_i(t)$ the position of an arbitrary ion in the CEL with respect to the center of the CEL and considering that all ions move with a velocity close to the mean velocity $v_i^{(l)}$, we obtain:
\begin{eqnarray}
\frac{d^2}{dt^2}\chi_i = -\frac{Z\,E_0(t)}{m_i\,l_e\,\gamma_i^3}\,\chi_i\,.
\end{eqnarray}
Therefore, if both $E_0(t)$ and $\gamma_i$ vary slowly on a time scale $\Omega^{-1}$, where $\Omega^2 = Z\,E_0(t)/(m_i\,l_e\,\gamma_i^3)$, the ions in the CEL have an harmonic motion around the mean velocity. From this, one can infer the dispersion in ion velocities of the accelerated bunch $\Delta v_i \sim l_e\,\Omega$ in the frame moving with the CEL. Correspondingly, the relative energy dispersion for non-relativistic ions scales as:
\begin{eqnarray}\label{eq_relatdisp_PSA}
\frac{\Delta\Epsilon_i}{\Epsilon_i} \propto \left(\frac{Z\,l_e\,a_0}{\Epsilon_i}\right)^{1/2}\,.
\end{eqnarray}
This scaling with $\Epsilon_i^{-1/2}$ ensures, with the small CEL thickness $l_e \ll 1$, the quasi-monoenergetic feature of the accelerated ion beam. In the ultra-relativistic regime, one would obtain $\Delta\Epsilon_i/\Epsilon_i \propto \Epsilon_i^{-5/2}$. However, and as it will be shown in numerical simulations (Sec.~\ref{sec4.2}), the reduction of radiation pressure in the target frame associated with its relativistic recoil does not allow for this acceleration regime to be maintained at ultra-relativistic velocities. Instead, acceleration will more and more evolve like the multi-stage process discussed in the next Sec.~\ref{sec3.2}.

Phase-stable acceleration of thin target thus opens an interesting path toward the creation of energetic quasi-monochromatic ion beams. Nevertheless, there is one restriction which was not mentioned in the original paper by Yan {\it et al.}~\cite{yan_PRL_2008} that we want to address now. In this specific regime of laser-target interaction, the electron bunch is compressed at the target rear-side. A large energy can thus be stored in the electrostatic field. Once the laser pulse is turned off, this energy is transferred back to the electrons, which start to quiver around the target, thus inducing its adiabatic expansion and in turn widening the ion energy spectrum.

To estimate the importance of this effect, we derive the energy stored in the electrostatic field (see also~\cite{macchi_PRL_2009}). Neglecting the contribution of the CEL due to its small thickness and considering that the electrostatic field varies linearly, $E_x(x) = E_0(t)\,x/d(t)$ for $0<x<d(t)$, the energy stored in the electrostatic field reads:
\begin{eqnarray}
\Epsilon_{es}(t) \sim \int_0^{d(t)} \frac{E_x^2(x)}{2}\,dx = \frac{a_0^2}{3}\,\gamma_i^4\,\left(1-v_i^{(l)}\right)^4\,d(t)\,.
\end{eqnarray}
For non-relativistic ion velocities this quantity simplifies to $\Epsilon_{es}(t) \sim n_{i0}\,d_0\,\Epsilon_i(t)/3$, suggesting that the energy stored in the electrostatic field is of the same order as the total ion kinetic energy. For such ion velocities, one should therefore expect a broadening of the ion energy distribution once the laser is turned off. For ultra-relativistic velocities however, $\Epsilon_{es}$ is found to remain much smaller than the total kinetic ion energy, and the effect of adiabatic expansion on energy dispersion is negligible.

These theoretical predictions are compared to numerical simulations in Sec.~\ref{sec4.2}.

\subsection{Multi-stage RPA of thicker targets}\label{sec3.2}

The acceleration mechanism for $\xi \gg 1$ can be described as a multi-stage process~\cite{zhang_POP_2007,klimo_PRSTAB_2008}, where the target undergoes successive hole-boring (HB) processes. For the sake of clarity, we first present the multi-stage process in the case of non-relativistic ion velocities. Then we generalize the procedure to the relativistic case.

\subsubsection{Non-relativistic ion velocities}

In a first stage, ion acceleration follows from the laser-driven HB of the immobile target. The laser acts on the target ions as a piston, moving deeper into the target with the velocity $v_{p0}$ and reflecting an increasing number of ions~\cite{naumova_PRL_2009,schlegel_POP_2009}. The piston velocity can be derived easily by considering the balance of radiation and electrostatic pressure in the frame comoving with the piston. If the laser field amplitude $a_0$ and the target density $n_{i0}$ are constant, $v_{p0}$ will not change in time. For non-relativistic ion velocities, we obtain $v_{p0} = a_0/\sqrt{2\,m_i\,n_{i0}}$.  During this stage, the ion velocity in the laboratory-frame ranges between $v_i^{(l)}=0$ (corresponding to ions having not been picked up by the laser piston) and $v_{i,1}^{(l)}=2\,v_{p0}$ (corresponding to ions that have been reflected once by the laser piston). Index $1$ in the ion velocity denotes the first acceleration stage. This first stage lasts up to the time $\tau_0=d_0/v_{p0}$ when the piston reaches the initial position of the back of the target. Ideally, at the end of this stage, all ions of the target have been accelerated to the velocity $v_{i,1}^{(l)}=2\,v_{p0}$ in the laboratory-frame and the whole target has been accelerated as a quasi-neutral bunch. At this point, we restrict ourselves to thin targets with thickness $d_0 \ll v_{p0}\,t_{p}$ (with $t_p$ the laser pulse duration), a necessary condition for the multi-stage process of RPA. Targets with larger thickness will only undergo HB.

To describe the second acceleration stage, i.e. for times $t>\tau_0$, we consider that the whole target is moving with the velocity $v_t^{(l)} = v_{i,1}^{(l)} = 2\,v_{p0}$ in the laboratory-frame. Then, in the frame moving with the target, ion acceleration proceeds in a way similar to laser-driven HB. During this stage, ion velocities range between $0$ and $2\,v_{p0}$ in the target-frame, which transforms to $v_t^{(l)} = 2\,v_{p0}$ and $v_{i,2}^{(l)} = v_t^{(l)}+2\,v_{p0}=4\,v_{p0}$ in the laboratory-frame.

Then, if $t_p > 2\,\tau_0$, a third acceleration stage starts during which the ion velocity in the laboratory frame ranges between $4\,v_{p0}$ and $6\,v_{p0}$. For sufficiently long laser pulse, this multi-stage process goes on so that at the $j^{th}$ stage, the ion velocity ranges between $v_{i,j-1}^{(l)}=2\,(j-1)\,v_{p0}$ and $v_{i,j}^{(l)}=2\,j\,v_{p0}$. Correspondingly, the ion energy ranges between $2\,(j-1)^2\,m_i\,v_{p0}^2$ and $2\,j^2\,m_i\,v_{p0}^2$. Generation of quasi-monoenergetic ion bunches thus requires the acceleration process to occur over many steps, i.e. over a time $t \gg \tau_0$. If this condition is satisfied, the ion energy and energy dispersion at a time $t \sim j\,\tau_0$ ($j \gg 1$) read:
\begin{eqnarray}
\Epsilon_i(t) &=& 2\,m_i\,v_{p0}^2\,\left(t/\tau_0\right)^2\,,\\
\Delta\Epsilon_i(t) &=& (2\,\tau_0/t)\,\Epsilon_i(t)\,.
\end{eqnarray}
The first equation corresponds to the ion energy evolution expressed by Eq.~(\ref{eq_ionenergy_nonrelat}), which was obtained using the macroscopic model in Sec.~\ref{sec2.1} in the limit of non-relativistic ion velocities. The second relation gives an information about the energy dispersion of the ion bunch. 

\subsubsection{Iterative procedure for relativistic ion velocities}

To extend the multi-stage model to higher ion velocities, relativistic effects such as the radiation pressure diminution on the target and dilation of the characteristic stage duration due to the target relativistic recoil have to be accounted for. 

The initial stage of ion acceleration is once more similar to the laser-driven HB of the immobile target. Accounting for relativistic effects, the piston velocity is $v_p = v_{p0}/(1+v_{p0})$. During this stage, ion velocities thus range between 0 (not yet reflected ions) and $v_{i,1}^{(l)}=2\,v_p/(1+v_p^2)$ (reflected ions). Correspondingly, we obtain the minimum and maximum ion energies during the first stage: $\Epsilon_{i,1}^{min}=0$ and $\Epsilon_{i,1}^{max}=(\gamma_i-1)\,m_i$, where $\gamma_i = (1-v_i^{(l)^2})^{-1/2}$. This stage ends when the piston reaches the initial position of the target rear-side, i.e. after a time $\tau^{(l)}_s = d_0/v_p$. In contrast to the classical limit, however, ions do not reach exactly twice the piston velocity. As a consequence, the target thickness, and consequently the ion density, have changed. At the end of this first acceleration stage, we have $d_1=v_{i,max}^{(t)}\,\tau_s^{(t)}-d_0$ and $n_{i,1}=n_{i0}\,d_0/d_1$.

Let us now consider the $j^{th}$ stage of the acceleration process ($j>1$). The target velocity (in the laboratory frame) is the velocity that ions have reached at the former $j-1$ stage : $v_t^{(l)} = v_{i,j-1}^{(l)}$. In the frame moving with the target, the laser radiation pressure is therefore reduced by the factor $\gamma_t^2\,(1-v_t^{(l)})^2$ [where $\gamma_t = (1-v_t^{(l) 2})^{-1/2}$]. The piston velocity in the target frame thus reads $v_p^{(t)} = v_p'/(1+v_p')$, where $v_p' = v_{p0}\,\big( n_{i0} / n_{i,j-1} \big)^{-1/2}\,\gamma_t\,\big(1-v_t^{(l)}\big)\,$. It follows that the reflected ions have a velocity $v_i^{(t)} = 2\,v_p^{(t)}/(1+v_p^{(t) 2})$ in the frame moving with the target, which transforms in $v_{i,j}^{(l)} = (v_i^{(t)}+v_t^{(l)})/(1+v_i^{(t)}\,v_t^{(l)})$ in the laboratory-frame. The duration of the stage in the target frame is easily computed as $\tau_s^{(t)}=d_{j-1}/v_p^{(t)}$, while one has to account for time dilation in the laboratory frame $\tau_s^{(l)}=\gamma_t\,\tau_s^{(t)}$. As for the target thickness and density at the end of the stage, they have to be recalculated in the target frame as $d_j = v_i^{(t)}\,\tau_s^{(t)}-d_{j-1}$ and $n_{i,j}=n_{i0}\,d_0/d_j$.

\begin{figure}
\begin{center} 
\includegraphics[width=12cm]{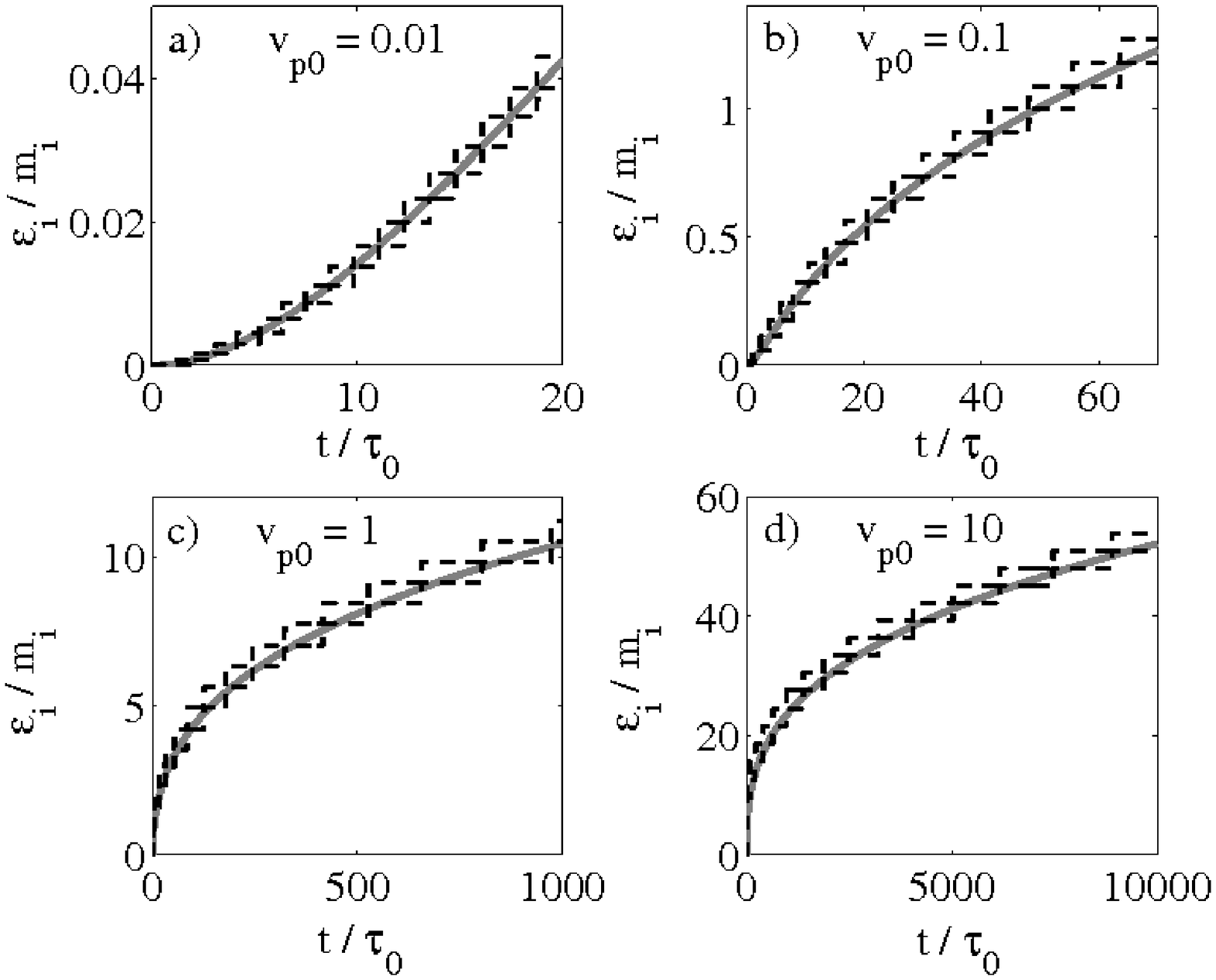}
\end{center}
\caption{Ion maximum and minimum energy predicted by the multi-stage model (dashed curves) and comparison to predictions from the macroscopic model (gray solid curves) for: a)~$v_{p0}=0.01$, b)~$v_{p0}=0.1$, c)~$v_{p0}=1$ and d)~$v_{p0}=10$.}
\label{fig_anl1}
\end{figure}

Following this procedure, we can compute the temporal evolution of the minimum and maximum ion energies for arbitrary values of the parameters $v_{p0}$ and $d_0$. The comparison of this multi-stage model with the macroscopic model of Sec.~\ref{sec2.1} is given in Fig.~\ref{fig_anl1} for several values of $v_{p0}$. Predictions from the multi-stage model match perfectly with results from Sec.~\ref{sec2.1} for $v_{p0}=0.01$, where the characteristic ion energy evolution $\propto (t/\tau_0)^2$ is recovered (Fig.~\ref{fig_anl1}a). For higher values of $v_{p0}$ (Fig.~\ref{fig_anl1}b-d), a still good agreement is found between the two models. A small discrepancy can however be observed in the ion mean energy\footnote{Our multi-stage model predicts a slightly higher ion energy than the macroscopic model. This is because, in our multi-stage model, we have a step-like decrease of the radiation pressure onto the target, while radiation pressure is continuously decreasing in the macroscopic model.}, but it remains small compared to the predicted energy dispersion. It must also be noted that the characteristic duration of an acceleration stage, for relativistic ion velocities, can be strongly dilated in the laboratory-frame, which has an important effect on  energy dispersion.
These analytical predictions are compared to numerical simulations in Sec.~\ref{sec4.3}.

%%%%%%%%%%%%%%%%%%%%%%%%%
% NUMERICAL SIMULATIONS %
%%%%%%%%%%%%%%%%%%%%%%%%%
\section{Numerical simulations}\label{sec4}

Numerical simulations of RPA of thin targets have been performed using the PIC code PICLS~\cite{sentoku_JCP_2008}. In order to make a direct comparison with our analytical model, only one-dimensional in space and three-dimensional in velocities (1D3V) simulations are presented. This choice is also justified as it has been shown that a quasi-1D geometry is required to avoid strong electron heating and partial transparency of the foil, which may prevent quasi-monoenergetic ion beam generation~\cite{klimo_PRSTAB_2008}. Our study thus provides us with necessary, albeit not sufficient, conditions for creating monoenergetic ion beams. 

In our simulations, a circularly polarized laser pulse is focused at normal incidence on a thin, fully ionized, Carbon target with density $n_{i0} = 25$ and $Z=6$. The target is located at a distance $2\,\lambda_L$ from the left boundary of the simulation box (the laser propagates from left to right). Both the incident laser field amplitude and the target thickness are varied to explore different regimes of RPA in thin foils. 

\subsection{Optimal target thickness}\label{sec4.1}

\begin{figure}
\begin{center} 
\includegraphics[width=6cm]{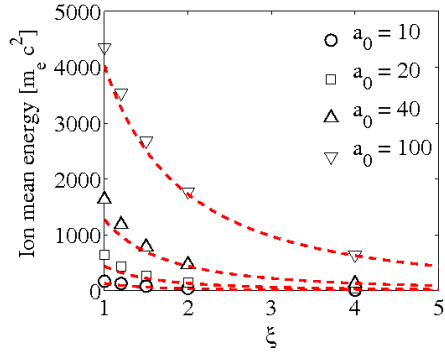}
\end{center}
\caption{Dependence of the ion mean energy on the normalized parameter $\xi$ (proportional to the target thickness) $\simeq 10\,\tau_L$ after the beginning of the interaction. The carbon target has density $Z\,n_{i0}=150$ and the laser field amplitude is $a_0=10$ ($\circ$), $a_0=20$ ($\Box$) and $a_0=40$ ($\vartriangle$). Red dashed lines correspond to predictions from the macroscopic (light-sail) model [solutions of Eq.~(\ref{eq_motion_foil})]. }
\label{fig_pic0}
\end{figure}

Figure~\ref{fig_pic0} shows the ion mean energy at an instant $t\simeq 10~\tau_L$ after the beginning of the interaction for $\xi \ge 1$. In the case $\xi<1$, the simulations indeed confirm that all electrons are removed from the target due to the strong radiation pressure, leading to the Coulomb explosion of the non-neutralized ion layer. For $\xi \ge 1$, one observes that the ion mean energy increases as the target thickness decreases: the lighter the target, the higher ion energy one can reach. Such a result was already discussed in Ref.~\cite{rykovanov_NJP_2008,macchi_PRL_2009} with the conclusion that the regime, $\xi \sim 1$, is the optimum case for high energy ion generation. 

\subsection{Ion acceleration in the phase-stable regime}\label{sec4.2}

\begin{figure}
\begin{center} 
\includegraphics[width=\columnwidth]{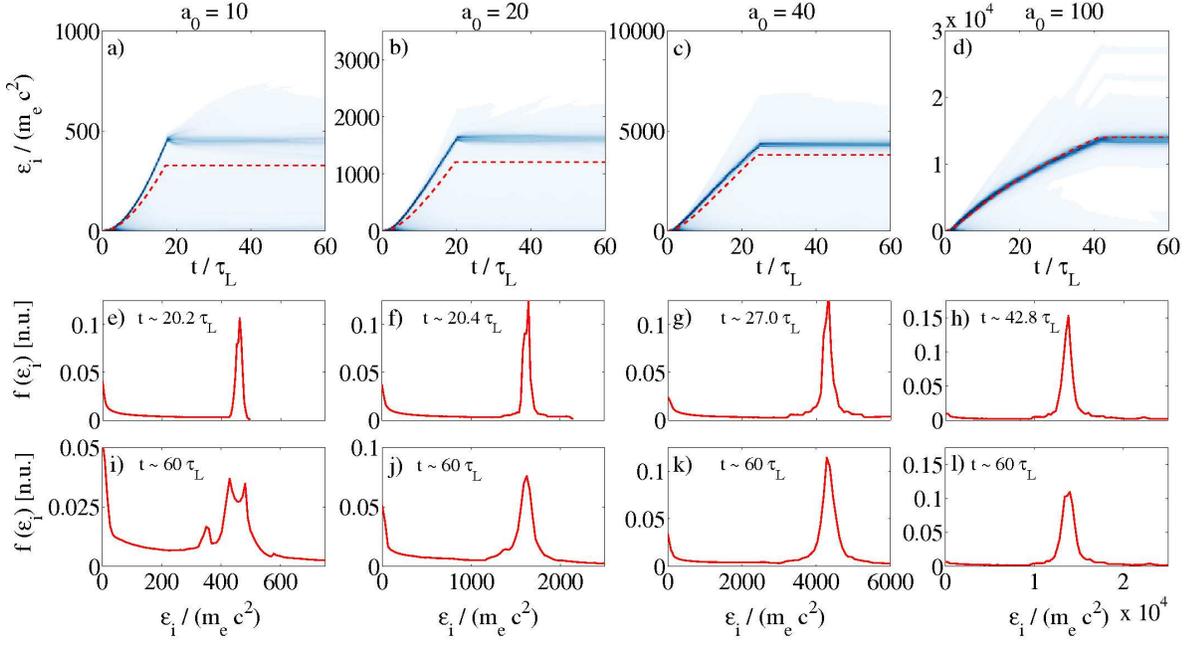}
\end{center}
\caption{a-d) Time-resolved ion energy spectra. e-h) Snapshot of the energy spectrum at the end of the laser-target interaction. i-l) Snapshot of the energy spectrum at the end of the simulation. The laser field amplitude is: a,e,i) $a_0 = 10$, b,f,j) $a_0=20$, c,g,k) $a_0=40$ and d,h,l) $a_0=100$. Red dashed lines show theoretical predictions from the macroscopic model.}
\label{fig_psa1}
\end{figure}

\begin{figure}
\begin{center} 
\includegraphics[width=7cm]{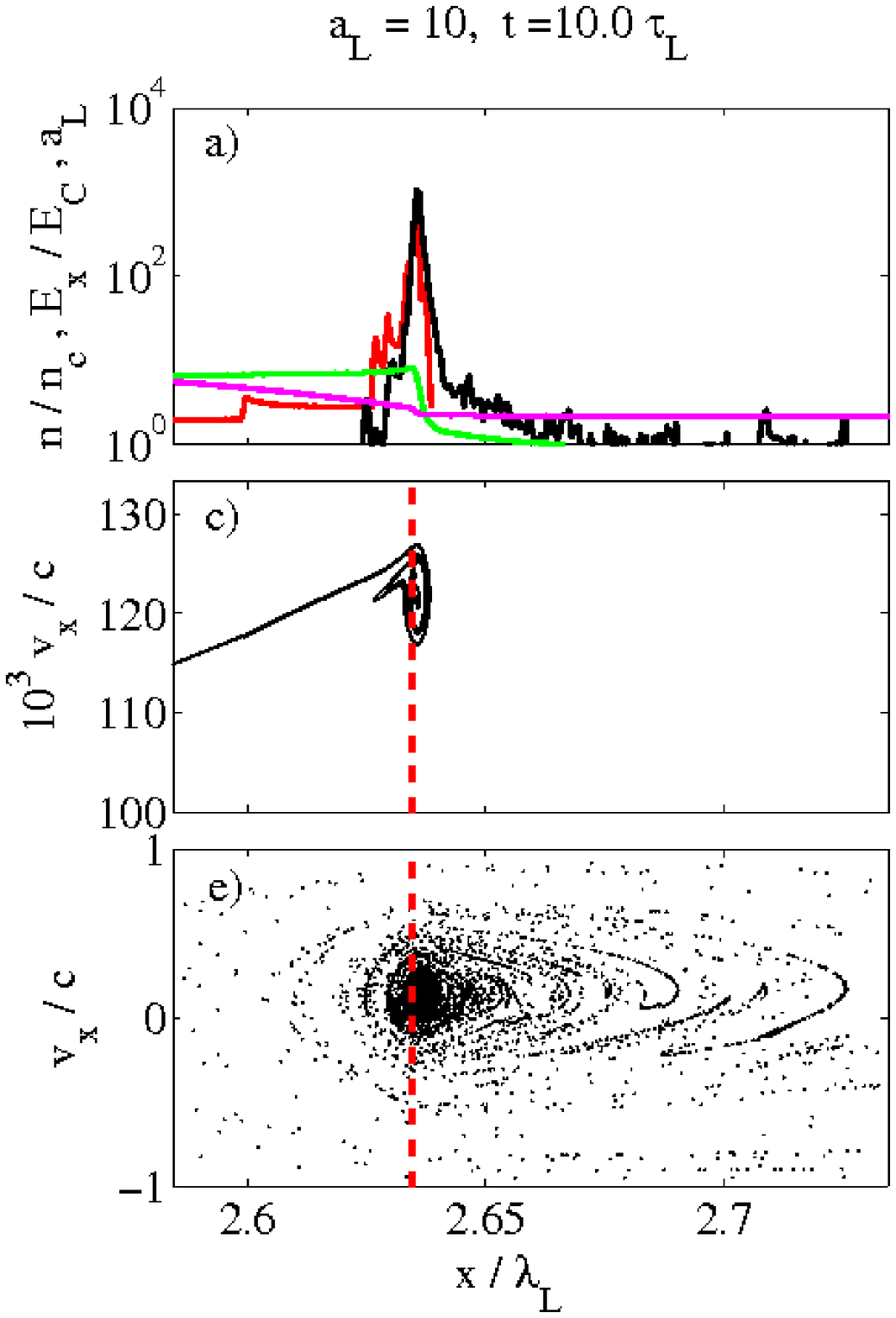}
\includegraphics[width=7cm]{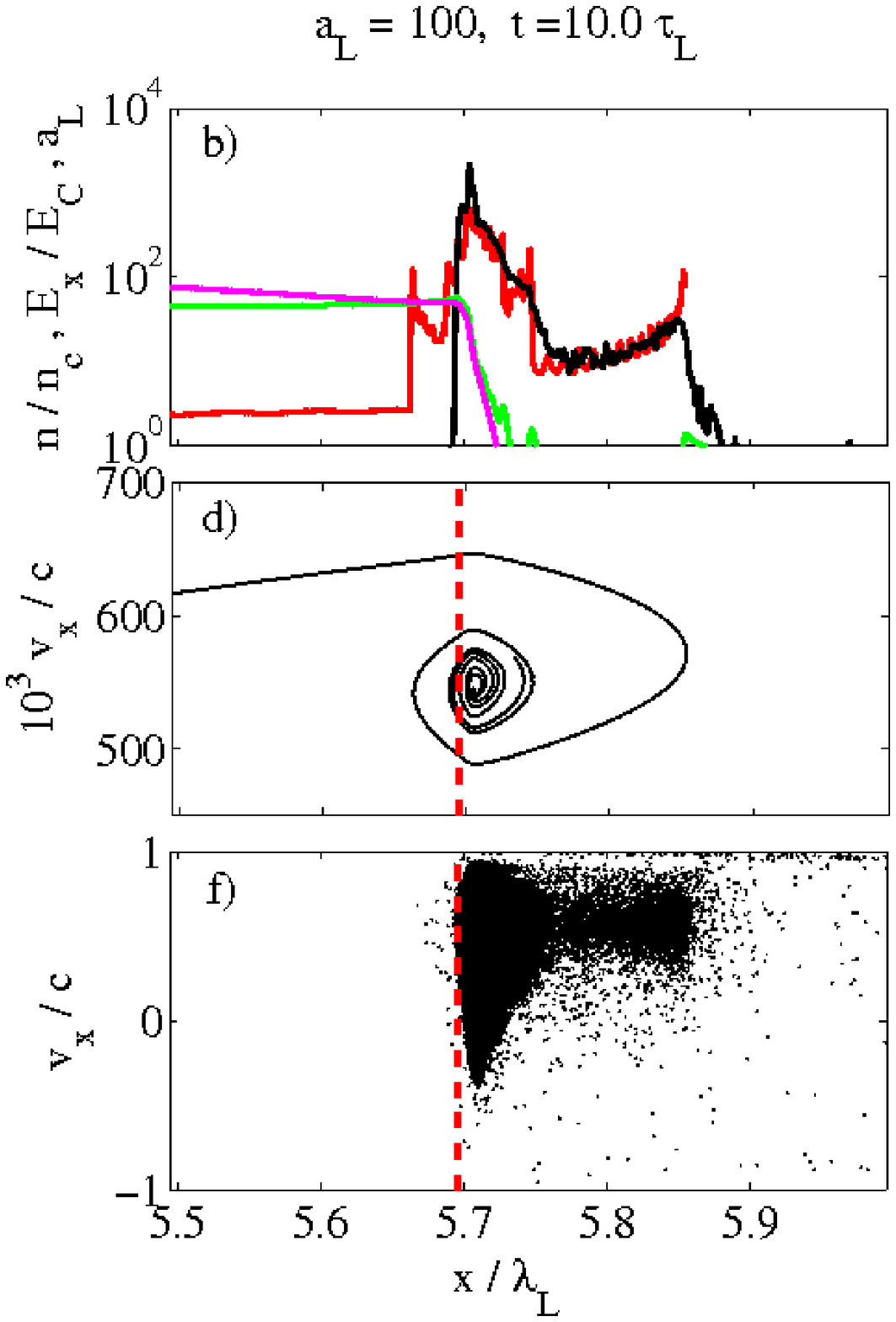}
\end{center}
\caption{Details of the target $10\,\tau_L$ after the beginning of the interaction for a,c,e)~$a_0=10$, and b,d,f)~$a_0 = 100$. a,b) The laser field amplitude is shown in magenta, the electrostatic field in green and the ion and electron densities in red and black, respectively. c,d) Ion distribution in phase-space. e,f) Electron distribution in phase-space. See also Movies~1 and~2 available from \url{www.pks.mpg.de/~grech/RPADE/mvie1.avi} [0.8~MB] and \url{www.pks.mpg.de/~grech/RPADE/mvie2.avi} [1.6~MB].}
\label{fig_psa2}
\end{figure}

Let us have a more detailed look into ion acceleration in the phase-stable regime $\xi \sim 1$. Figure~\ref{fig_psa1} presents the time-resolved energy spectra obtained in simulations for different incident laser field amplitudes $a_0 = 10-100$ and $\xi \sim 1$ (the target thickness is adjusted for each laser amplitude). Theoretical predictions for the ion mean energy from the macroscopic model (Sec.~\ref{sec2.1}) are superimposed on the numerical results. A rather good agreement is found between theory and simulation. However, while for $a_0=100$ the ion mean energy evolves exactly as predicted by the macroscopic model, numerical simulations at lower field amplitudes ($a_0 = 10 - 20$) show higher ion energies than estimated analytically. The reason of this discrepancy can be found in Fig.~\ref{fig_psa2} and the corresponding movies, where details of the temporal evolution of the target structure during the acceleration process are presented for $a_0=10$ and $a_0=100$. For the lower laser field amplitude, $a_0=10$, the CEL is not totally opaque to the laser field, which is partly transmitted (the foil transmittance is here $T \simeq 20~\%$). The ponderomotive force on the target rear-side is thus non zero and some electrons can escape from the target into the vacuum behind. This widens the electrostatic field distribution and increases its average value at the target rear-side thus leading to the observed increase of ion energy. In contrast, for $a_0=100$, the target is partly transparent only during a short time at the beginning of the interaction. Once the target has reached a relativistic velocity, the radiation pressure in the target frame is lower and electrons remain confined in the target. On such long times, the CEL does not actually stay at the target rear-side and ion acceleration becomes more similar to multi-stage acceleration. 

\begin{figure}
\begin{center} 
\includegraphics[width=6cm]{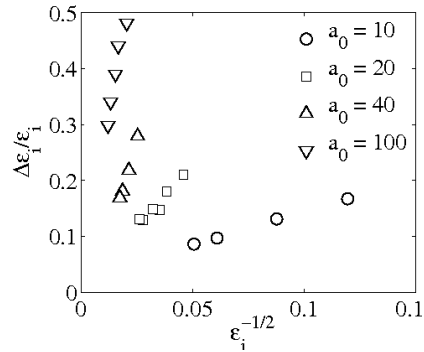}
\end{center}
\caption{Relative energy dispersion during the phase-stable acceleration as a function of the ion mean energy. For $a_0=10$ ($\circ$), $a_0=20$ ($\Box$), $a_0=40$ ($\vartriangle$) and $a_0=100$ ($\triangledown$).}
\label{fig_psa3}
\end{figure}

Figure~\ref{fig_psa1} also reveals the quasi-monoenergetic ion distribution during the acceleration process. This is underlined in the different panels of Fig.~\ref{fig_psa1} presenting the ion energy at the end of the interaction process (panels~\ref{fig_psa1}e-h), as well as in Fig.~\ref{fig_psa3} which shows the relative ion energy dispersion as a function of the ion mean energy for $a_0=10-100$. Figure~\ref{fig_psa3} also confirms the characteristic dependency of the relative energy dispersion, which is proportional to $\Epsilon_i^{-1/2}$ as predicted by Eq.~(\ref{eq_relatdisp_PSA}). Equation~(\ref{eq_relatdisp_PSA}) also predicts that $\Delta\Epsilon_i/\Epsilon_i$ should scale as $\sqrt{l_e\,a_0}$, and one could naively expect for a given ion mean energy, that $\Delta\Epsilon_i/\Epsilon_i$ scales as the square-root of the laser field amplitude $a_0$. Figure~\ref{fig_psa3} however shows that the dependence on $a_0$ is stronger. This is because the CEL thickness $l_e$ actually increases with $a_0$ for a fixed value of $\xi \propto d_0/a_0$. 

Let us now discuss the fraction $f_i$ of ions in the monoenergetic peak. A naive estimate can be derived from the semi-microscopic model presented in Sec.~\ref{sec3.1} by considering that only ions initially located in the CEL participate in the phase-stable acceleration so that $f_i \sim 1-\xi^{-1}$. This estimate would suggest that only a very small fraction of the target ions participate in the acceleration process. Simulations however show quite the contrary: $f_i$ ranges between $0.45$ and $0.67$ for $a_0=10-100$. There are two reasons for such a high number of accelerated ions. (i) While in the model presented in Sec.~\ref{sec3.1} the CEL thickness shrinks to $0$ as $\xi \rightarrow 1$, the electron pressure in the CEL actually increases during the compression by the laser pulse and prevents its collapse. As a result, the CEL thickness is much larger than expected in the model and so is the fraction of accelerated ions. (ii) Furthermore, some ions initially located outside of the CEL can still be injected (after some time) in the CEL and thereafter participate in the phase-stable acceleration. While creation of the CEL is almost instantaneous (because of electrons relativistic velocities, it occurs on a characteristic time $\sim d_0/c \ll 1$), ions react to the strong electrostatic field on a longer time scale of the order of the inverse ion plasma frequency. A self-maintained structure made of both the CEL and ions is formed and it is this structure which is accelerated in a phase-stable way. The correct modelling of this accelerating structure is very challenging as it would require to describe self-consistently the ion and electron motions. While this has been done for ion acceleration in the HB regime in Ref.~\cite{schlegel_POP_2009} by developing a quasi-stationary model, this is particularly difficult under current conditions as the quasi-stationary hypothesis does not hold. What is also clearly highlighted in the two movies is that some ions, which are not initially located inside the accelerating structure, can be injected into it after some time. Ions located outside the CEL indeed 'see' a constant accelerating field and may actually catch up with the CEL after some time. For non relativistic ions, one can easily estimate that only ions initially located at a position $x_{i0}>d_0/2$ can reach the CEL and their fraction $f_i$ cannot exceed $50~\%$. Accounting for relativistic effects allows for a larger fraction of reinjected ions (which increases in time, according to simulations, up to $67~\%$ for $a_0=100$). 

\begin{figure}
\begin{center} 
\includegraphics[width=8cm]{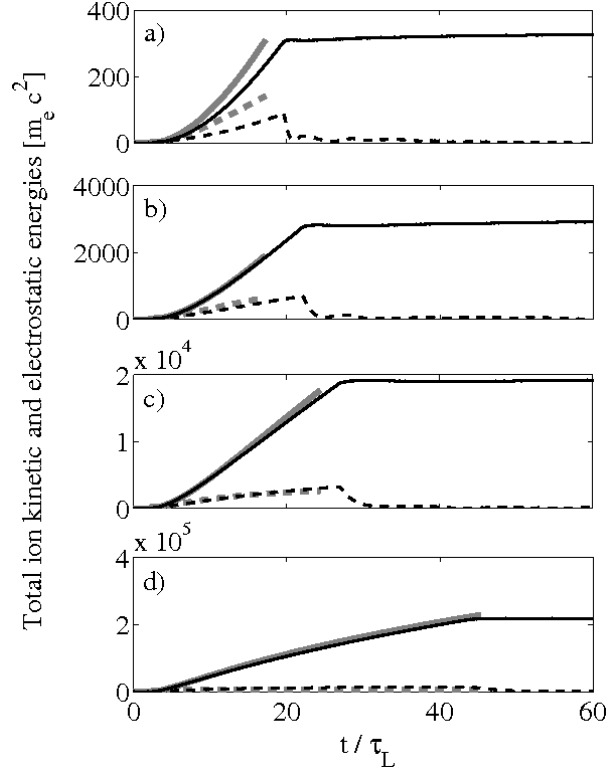}
\end{center}
\caption{Temporal evolution of the total kinetic energy of ions in the monoenergetic peak (solid curves) and the energy stored in the electrostatic field (dashed curves) for: a)~$a_0=10$, b)~$a_0=20$, c)~$a_0=40$ and d)~$a_0=100$. Gray lines show theoretical predictions from Sec.~\ref{sec3.1}.}
\label{fig_psa4}
\end{figure}

Finally, we want to point out that, once the laser is turned-off, the energy stored in the electrostatic field goes back to electrons which start quivering around the accelerated ions thus driving its adiabatic expansion. If the electrostatic energy is of the same order of magnitude as the total ion energy, this can strongly enhance the final ion energy dispersion. To investigate this effect in-depth, we have plotted the total energy of ions in the monoenergetic peak and the energy stored in the electrostatic field as a function of time for $a_0=10 - 100$ in Fig.~\ref{fig_psa4}. For small laser amplitudes $a_0=10 - 20$ and correspondingly non-relativistic ion energies, a non-negligible fraction of the energy ($\simeq 30\,\%$) is stored in the electrostatic field. As a consequence, the corresponding ion energy spectra (Figs.~\ref{fig_psa1}i,j) are considerably wider at the end of the simulation. On the contrary, and as predicted in Sec.~\ref{sec3.1}, this effect is negligible for sufficiently large ion energies. For $a_0=40 - 100$, most of the energy is stored in the accelerated ion bunch and no-enhanced ion energy dispersion is observed (Figs.~\ref{fig_psa1}k,l).

\begin{figure}
\begin{center} 
\includegraphics[width=\columnwidth]{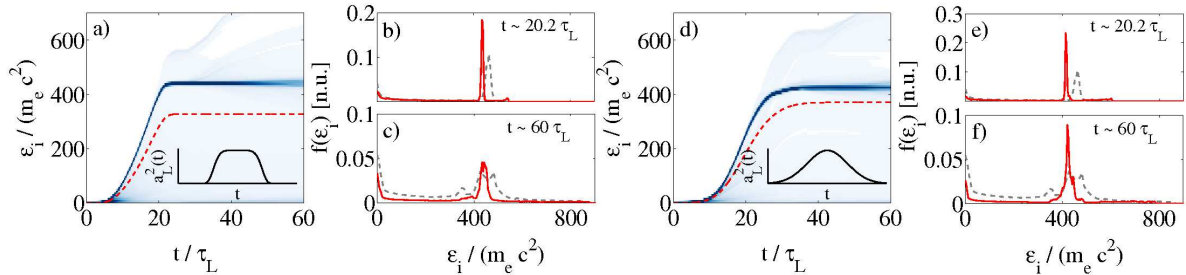}
\end{center}
\caption{Same as Fig.~\ref{fig_psa1} for $\xi=1$, $a_0=10$ and two different laser intensity profiles: a,b,c) for a $6^{th}$-order hyper-Gaussian profile, d,e,f) for a Gaussian profile. Red dashed lines in panels a) and d) show theoretical predictions from the light-sail model [solutions of Eq.~(\ref{eq_motion_foil})]. Gray dashed lines in panels b,c,e,f) show the energy spectra obtained using a rectangular laser pulse profile.}
\label{fig_psa5}
\end{figure}

In addition, we notice that this effect can also be mitigated at low ion energy by using a more sophisticated laser pulse temporal profile, e.g. by considering Gaussian or hyper-Gaussian pulses. Figure~\ref{fig_psa5} shows the ion energy spectra obtained using a $6^{th}$-order hyper-Gaussian or Gaussian laser pulse with similar maximum field amplitude $a_0=10$, fluence and full-width at half-maximum. The enhanced energy dispersion due to the long-time behavior of electrons is strongly mitigated as the pulse is slowly turned off.

\subsection{Ion acceleration in the multi-stage regime}\label{sec4.3}

\begin{figure}
\begin{center} 
\includegraphics[width=\columnwidth]{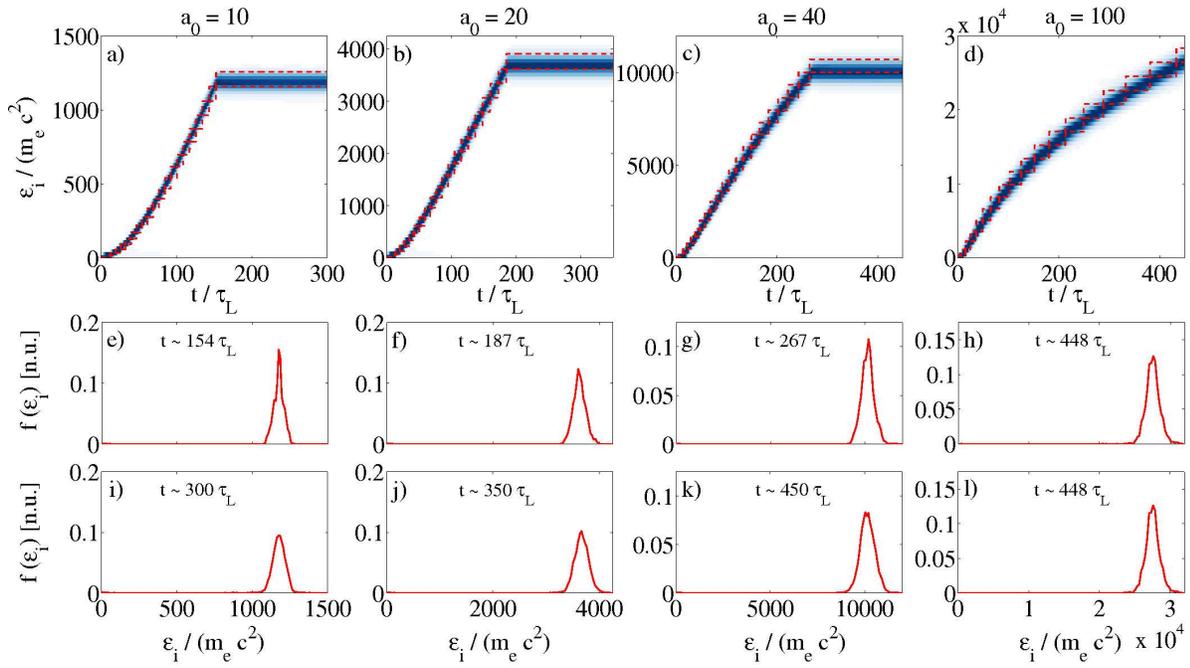}
\end{center}
\caption{Same as Fig.~\ref{fig_psa1} for $\xi=4$. Red dashed lines in panels a-d) show theoretical predictions from the iterative model (Sec.~\ref{sec3.2}).}
\label{fig_msa1}
\end{figure}

PIC simulations in the regime of multi-stage acceleration ($\xi \gg 1$) are now discussed. Figure~\ref{fig_msa1} shows the time resolved ion energy spectrum extracted from simulations with $\xi = 4$ and different laser field amplitudes ($a_0=10-100$). In these simulations, the laser temporal profile is constant over a duration corresponding to $20\,t_0$ (we recall that $t_0=d_0/v_{p0}$ is the characteristic duration of an acceleration stage). Theoretical predictions from the model developed in Sec.~\ref{sec3.2} are superimposed to the numerical results. A rather good agreement on the minimum and maximum ion energies as a function of time is found between theory and simulations: our model allows to correctly predict both the ion mean energy and energy dispersion. For cases with $a_0=40$ and $a_0=100$, ions quickly reach relativistic energies and the dilation in the laboratory frame of the characteristic stage duration becomes obvious. Also, the duration of the laser-target interaction (and therefore of the acceleration process) increases as the foil reaches larger velocities (Fig.~\ref{fig_msa1}). 

\begin{figure}
\begin{center} 
\includegraphics[width=7cm]{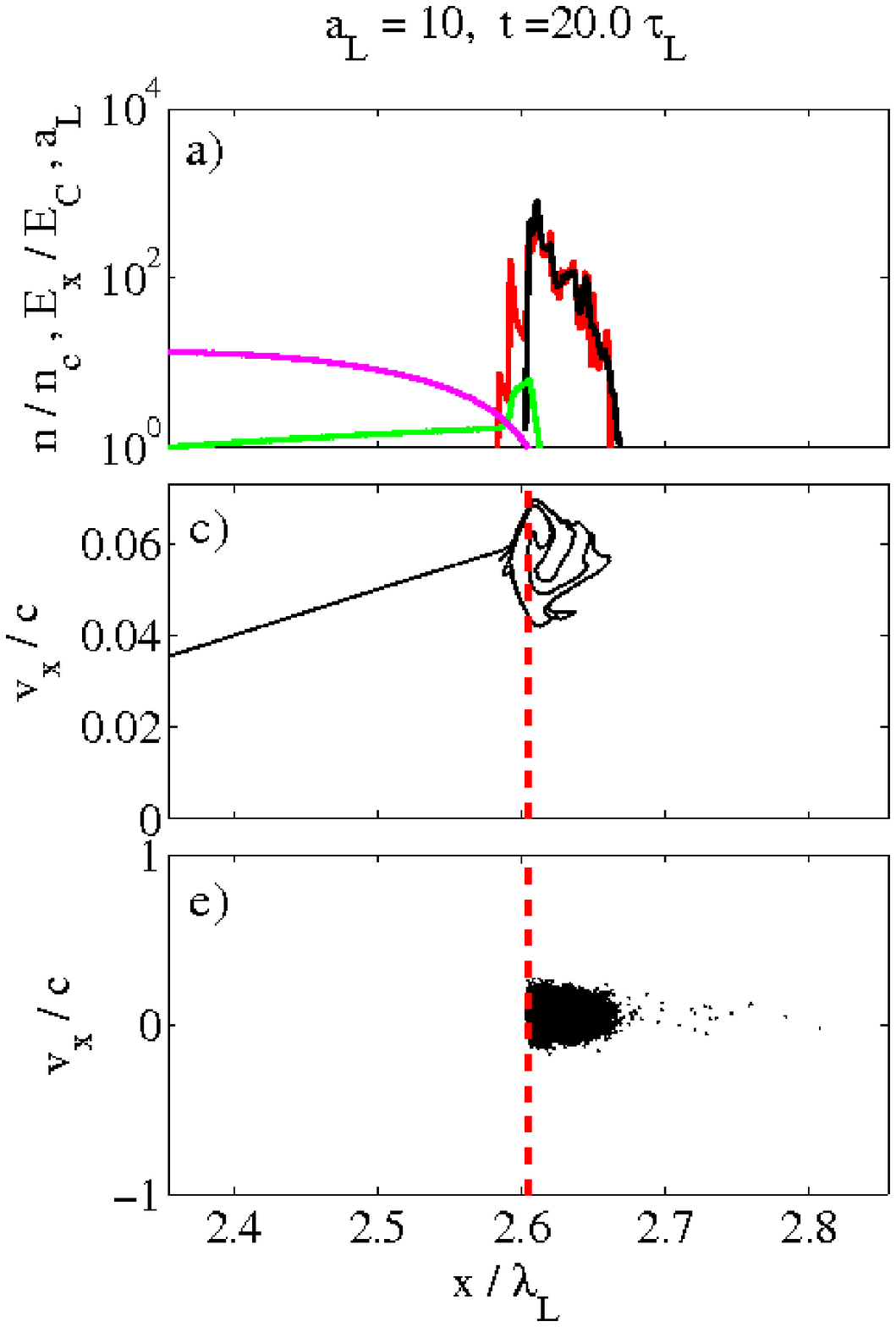}
\includegraphics[width=7cm]{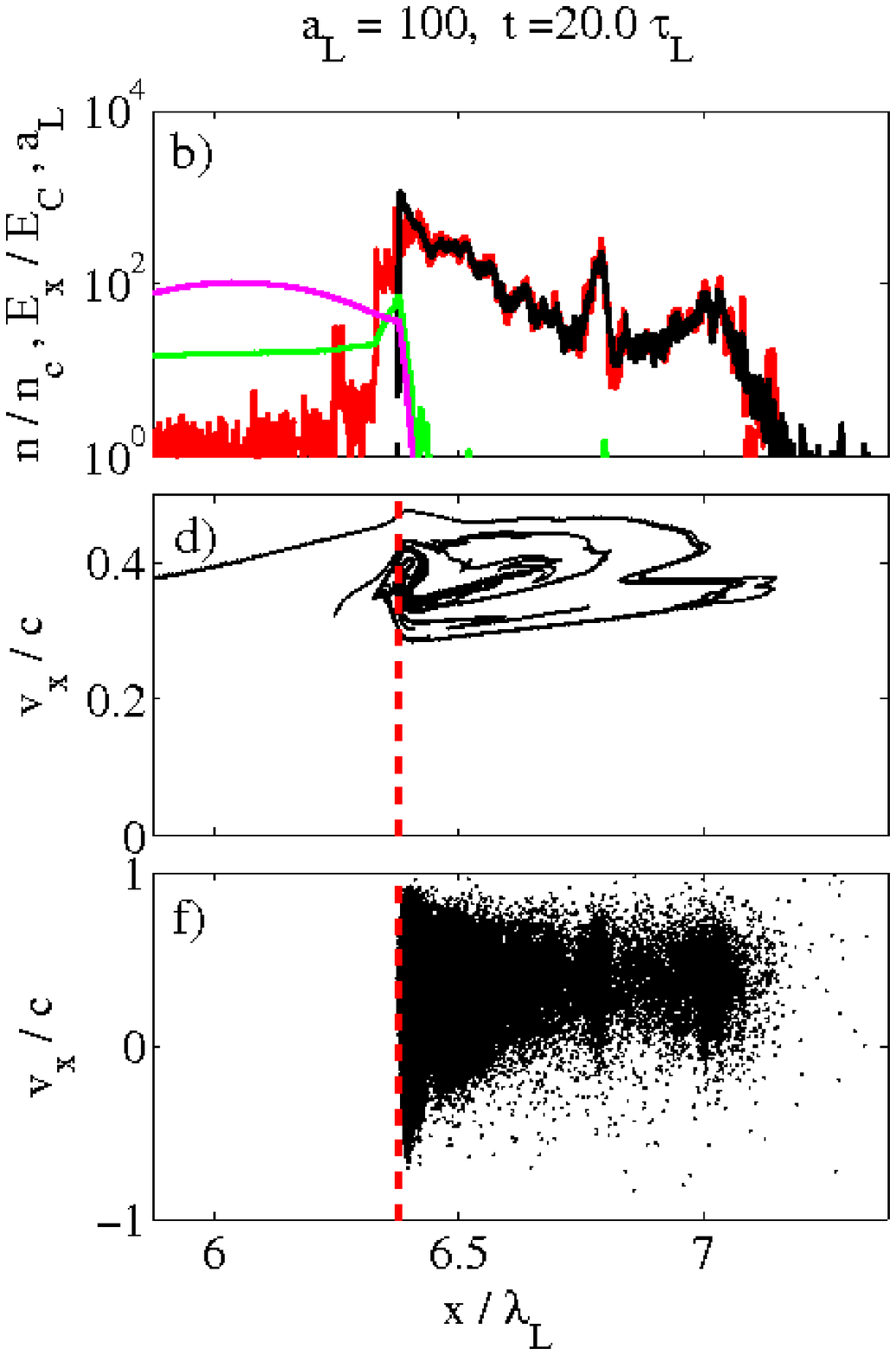}
\end{center}
\caption{Details of the target $20\,\tau_L$ after the beginning of the interaction for a,c,e)~$a_0=10$, and b,d,f)~$a_0 = 100$. a,b) The laser field amplitude is shown in magenta, the electrostatic field in green and the ion and electron densities in red and black, respectively. c,d) Ion distribution in phase-space. e,f) Electron distribution in phase-space. See also Movies~3 and~4 available from \url{www.pks.mpg.de/~grech/RPADE/mvie3.avi} [1.6~MB] and \url{www.pks.mpg.de/~grech/RPADE/mvie4.avi} [3.1~MB].}
\label{fig_msa1bis}
\end{figure}

\begin{figure}
\begin{center} 
\includegraphics[width=16cm]{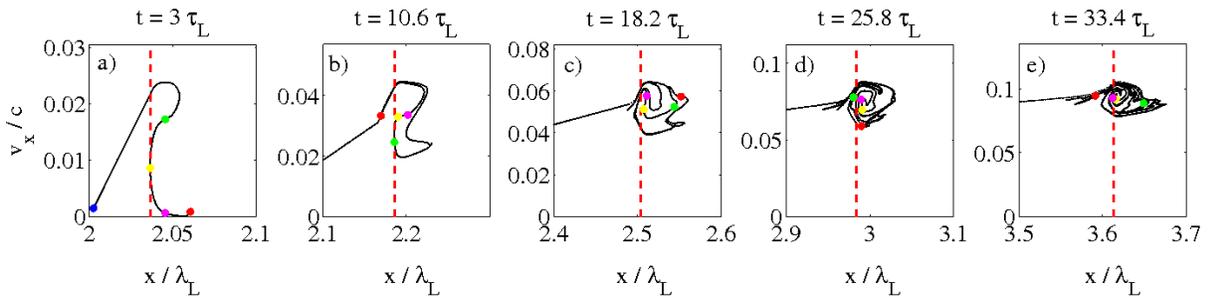}
\end{center}
\caption{Ion phase-space at different times after the beginning of interaction: a) $3~\tau_L$, b) $10.6~\tau_L$, c) $18.2~\tau_L$, d) $25.8\tau_L$ and e) $33.4~\tau_L$. Each time corresponds to a different acceleration stage. In this simulation, $a_0=10$ and $\xi=4.0$. The color dots follow test ions during the acceleration process. The vertical dashed line shows the position of the laser piston (position of the maximum electrostatic field).}
\label{fig_msa2}
\end{figure}

Additional details on the target structure as well as the ion and electron phase-space distributions during the acceleration process are given in Figs.~\ref{fig_msa1bis} and~\ref{fig_msa2} and in the corresponding movies. The transition between successive acceleration stages is clearly visible in the movies. For the case $a_0=10$ for instance, the first HB stage terminates at $t \simeq 16~\tau_L$. As ions are reflected again and again by the laser piston, their distribution in phase-space becomes more and more complex in contrast to what was observed during phase-stable acceleration (compare for instance Figs.~\ref{fig_msa1bis}c,d and Figs.~\ref{fig_psa2}c,d).

Furthermore, the ion energy spectra at the end of the simulations (Figs.~\ref{fig_msa1}i-l) are quite similar to those obtained at the end of the laser-target interaction (Figs.~\ref{fig_msa1}e-h). In contrast to phase-stable acceleration, the ion energy distribution here is not sensitive to the late time behavior of the electrons. This is due to the small fraction of laser energy which is stored in the electrostatic field at $\xi \gg 1$.

This excellent control of the ion beam spectrum as well as the large fraction of accelerated ions ($f_i$ exceeds $90~\%$ in these simulations) make this regime of acceleration especially attractive for high-quality ion beam generation. Numerical simulations also suggest that RPA proceeds in the multi-stage regime as soon as $\xi \gtrsim 2$, which makes this robust mechanism more likely to be observed in experiments than phase-stable acceleration.

Finally, as RPA in the multi-stage regime follows from successive HB of the target, one may suggest an additional source of energy dispersion for large laser field amplitudes and/or rather thick targets. Recent studies have indeed underlined the non-stationary behavior of the laser piston at high laser intensities giving rise to the so-called piston oscillations~\cite{schlegel_POP_2009,eliasson_NJP_2009}. This phenomenon implies large-amplitude (typically $\simeq 30~\%$) oscillations of the maximum electrostatic field in the laser piston leading to an enhanced energy dispersion of ions reflected during the HB process. While the origin of these oscillations is yet not fully understood, some of their characteristic features are known: (i) their characteristic period is of the order of the inverse ion plasma frequency ($\omega_{pi} = \sqrt{Z^2\,n_{i0}/m_i}$ in our normalized units), (ii) they appear after a characteristic time $t_s$ which is shorter for larger laser field amplitudes and/or target densities. For instance, PIC simulations of HB of a thick carbon foil with ion density $n_{i0} = 25$ by a CP laser pulse with $a_0=40$ indicate that oscillations in the electrostatic field with a characteristic period $\simeq 1.4~\tau_L$ and $\simeq 35~\%$ amplitude with respect to the maximum field strength occur after a time $t_s \simeq 5~\tau_L$. Increasing the laser field amplitude to $a_0=100$ does not change the relative amplitude of the oscillations or their period but shortens the characteristic time of their appearance to $t_s \simeq 3.7~\tau_L$.

\begin{figure}
\begin{center} 
\includegraphics[width=8cm]{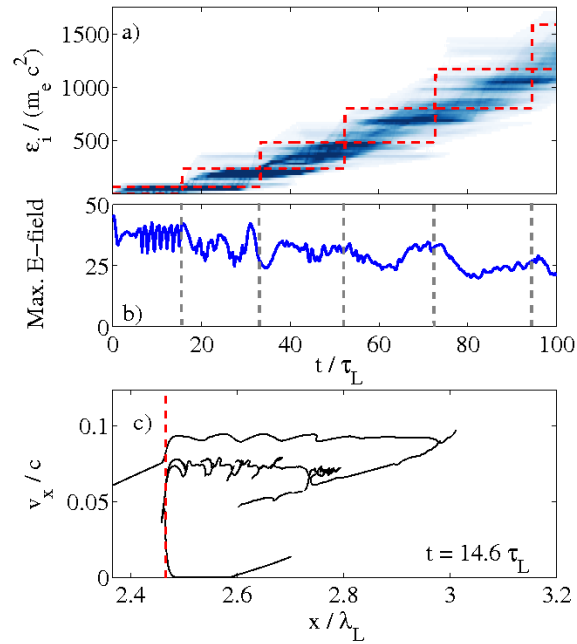}
\end{center}
\caption{a) Time-resolved ion energy spectra. The dashed lines show theoretical prediction from the multi-stage model. b) Temporal evolution of the maximum electrostatic field. Vertical gray lines indicate the end of the different acceleration stages as predicted from the multistage model. c) Ion phase-space $14.6~\tau_L$ after the beginning of the interaction. The vertical red line shows the position of the laser piston (where the electrostatic field is maximum).}
\label{fig_msa3}
\end{figure}

One could therefore fear that these piston oscillations widen the energy spectrum during the multi-stage acceleration. Clearly, this is not the case in the simulations presented in Fig.~\ref{fig_msa1}, where the energy dispersion is well described by our multi-stage model (which does not account for the piston oscillations). However, for these simulations, the characteristic duration of an acceleration stage $t_0 \sim d_0/v_{p0} \simeq 6~\tau_L$ is rather short and the piston oscillations have scarcely the time to develop. Therefore, we have performed a simulation with a thicker target ($\xi=10$ and $a_0=40$). Results from this simulation are presented in Fig.~\ref{fig_msa3}. During the first acceleration stage, for $t = 0 - 16~\tau_L$, oscillations in the maximum electrostatic field are clearly visible after a time $t \simeq 5~\tau_L$ (Fig.~\ref{fig_msa3}b) and result in a rather complex ion phase-space distribution (Fig.~\ref{fig_msa3}c as compared to Fig.~\ref{fig_msa2}a). Nevertheless, these regular oscillations disappear at the end of the first stage ($t > 16~\tau_L$), which we attribute to the already perturbed target configuration in the second acceleration stage. There are still some large variations in the maximum value of the electrostatic field, but these variations are generic to all realistic simulations we have performed in the multi-stage regime. They follow more from the global ion dynamics in the target than from the piston oscillations themselves. Hence, these results suggest that the so-called piston oscillations are not a concern for the control of ion energy distributions during RPA in thin foils.

%%%%%%%%%%%%%%%%%%%%%
%    CONCLUSIONS    %
%%%%%%%%%%%%%%%%%%%%%
\section{Conclusion}\label{sec5}

A detailed study of ion energy dispersion in RPA ion beams has been presented using both analytical modelling and 1D3V PIC simulations. The description proposed here allows for a greater insight in the details of RPA of thin foils than that available from the standard macroscopic light-sail model. In particular, it provides us with necessary conditions for quasi-monoenergetic ion beam generation.

Two RPA regimes are identified depending on the dimensionless parameter $\xi$ which determines, for a given laser field amplitude, the target thickness. For both regimes, the models we have developed allow to recover the ion energy temporal evolution obtained by considering efficient momentum transfer from the laser photons to the target ions (the usual light-sail model). By accounting for the target structure during the acceleration process, we have gained a deeper insight into the ion energy dispersion. For $\xi \sim 1$ (thin targets), RPA proceeds in the phase-stable regime introduced in Ref.~\cite{yan_PRL_2008}. Two sources of energy dispersion have been identified in this regime: the electric field inhomogeneity in the accelerating structure, and the adiabatic foil expansion due to the late time electron behavior. This later process is mainly important for low energy ions and its effect can be mitigated by using smooth temporal laser profiles. For $\xi \gtrsim 2$ (thicker targets), ion acceleration proceeds in the multi-stage regime originally discussed in Refs.~\cite{zhang_POP_2007} and~~\cite{klimo_PRSTAB_2008}, and for which we have developed a relativistic model. Ion energy dispersion in this regime is mainly determined by the number of acceleration stages. Small energy dispersion can thus be achieved by using long enough laser pulses. 

Hence, this work suggests that using moderately intense (and long) laser pulses is preferable for monoenergetic ion beam generation. The lower limit on the laser intensity actually follows from the need to accelerate the ions to the desired energy before the target escapes from the laser focal volume. As for the optimal target thickness, the thinner the target is, the higher ion energies one can reach. However, maintaining the target integrity in the regime $\xi \rightarrow 1$ might be experimentally difficult. Any nonuniformity in the laser intensity profile may lead to the removal of electrons and the resulting Coulomb explosion of the target. Also, Rayleigh-Taylor like instabilities, which have been observed in 2D simulations~\cite{klimo_PRSTAB_2008}, may be more detrimental for very thin targets. Since simulations suggest that RPA occurs in the multi-stage regime as soon as $\xi \gtrsim 2$, phase-stable acceleration may thus be difficult to achieve in experiments. We therefore expect multi-stage acceleration to be the practically relevant acceleration mechanism.

\section*{Acknowledgments}

The authors are grateful to E.~d'Humi\`{e}res and Y.~Sentoku for providing them with the code PICLS.

%%%%%%%%%%%%%%%%
% BIBLIOGRAPHY %
%%%%%%%%%%%%%%%%

%%%%%%%%%%%%%%%%
%   FIGURES    %
%%%%%%%%%%%%%%%%
\newpage

\end{document}